\def\rfr#1{eq. (\ref{#1})}

\def\virg#1{``#1''}

\def\eqi{\begin{equation}}
\def\eqf{\end{equation}}
\def\eqia{\begin{eqnarray}}
\def\eqfa{\end{eqnarray}}
\def\rp#1#2{{#1\over#2}} \def\lb#1{\label{#1}}

\documentclass[11pt]{article}
\usepackage{url}
\usepackage{amsmath,amsthm,amscd,amssymb}
\usepackage{latexsym,w-greek}
\usepackage{graphicx,epsfig}
\usepackage{hyperref}
\RequirePackage{color}

\begin{document}

\title{Constraints from orbital motions around the Earth of the environmental fifth-force hypothesis for the OPERA superluminal neutrino phenomenology}

\author{L. Iorio \\ Ministero dell'Istruzione, dell'Universit\`{a} e della Ricerca (M.I.U.R.)-Istruzione \\ Fellow of the Royal Astronomical Society (F.R.A.S.) \\
 International Institute for Theoretical Physics and
Advanced Mathematics Einstein-Galilei \\ Permanent address: Viale Unit$\grave{\rm a}$ di Italia 68
70125 Bari (BA), Italy \\ email: lorenzo.iorio@libero.it}

\maketitle

\begin{abstract}
It has been recently suggested by Dvali and Vikman that the superluminal neutrino phenomenology of the OPERA experiment may be due to an environmental feature of the Earth, naturally yielding a long-range fifth force of gravitational origin whose coupling with the neutrino is set by the scale $M_{\ast}$, in units of reduced Planck mass. Its characteristic length $\uplambda$ should not be smaller than one Earth's radius $ R_{\rm e}$, while its upper bound is expected to be slightly smaller than the Earth-Moon distance (60 $ R_{\rm e}$). We analytically work out some orbital effects of a Yukawa-type fifth force for a test particle moving in the modified field of a central body. Our results are quite general since they  are not restricted to any particular size of  $\uplambda$; moreover, they are valid for an arbitrary orbital configuration of the particle, i.e. for any value of its eccentricity $e$. We find that the dimensionless strength coupling parameter $\upalpha$ is constrained to $\left|\upalpha\right|\lesssim 1\times 10^{-10}-4\times 10^{-9}$  for $1\  R_{\rm e}\leq \uplambda \leq 10\ R_{\rm e}$ by the laser data of the Earth's artificial satellite LAGEOS II, corresponding to $M_{\ast} \textcolor{black}{\gtrsim} 4\times 10^9 -1.6\times 10^{10}$. The Moon perigee allows to obtain
$\left|\upalpha\right| \lesssim 3\times 10^{-11}$ for the Earth-Moon pair in the range $15\  R_{\rm e}\lesssim \uplambda \lesssim 60\  R_{\rm e}$, which translates as $M_{\ast} \textcolor{black}{\gtrsim} 3\times 10^{10} - 4.5\times 10^{10}$. Our results are \textcolor{black}{neither} necessarily limited to the superluminal OPERA scenario \textcolor{black}{nor to the Dvali-Vikman model, in which it is $M_{\ast}\sim 10^{-6}$ at $\uplambda\sim 1\ R_{\rm e}$, in contrast with our bounds}: they generally extend to any theoretical scenario implying a fifth-force of Yukawa-type.
\end{abstract}

\centerline
{PACS:  04.50.Kd; 96.30.-t; 95.10.Eg; 14.60.Lm; 14.60.St; 13.15.+g}

\section{Introduction}
The  measured arrival times
\textcolor{black}{\eqi \delta t = {57.8\pm \left.7.8\right.|_{\rm stat}}\left._{-5.9}^{+ 8.3}\right|_{\rm syst}  \ {\rm ns}\sim 58\pm 13\ {\rm ns}\lb{tempis}\eqf}
of the muon neutrinos ($\nu_{\mu}$) in the OPERA experiment \cite{opera} after a travel along a known baseline distance $d=730$ km from the CERN CNGS beam to the Gran Sasso Laboratory\footnote{See also the material pertaining to the conference held at CERN on September 2011 here: http://indico.cern.ch/conferenceDisplay.py?confId=155620.} (LNGS)  have been interpreted in terms of a property pertaining the motion of the neutrinos themselves. Indeed,  it was suggested that their speed $v_{\nu_{\mu}}$ may have overcome the speed of light in vacuum. Stated differently,
by defining the dimensionless quantity \eqi\xi\doteq \rp{v_{\nu_{\mu}}-c}{c},\lb{definizione} \eqf
it would be  \cite{opera}
\eqi \xi_{\rm meas}>0 \lb{misura}\eqf
at a statistical significant level.
The  result \cite{opera} \textcolor{black}{\eqi \xi = \left({2.37 \pm \left. 0.32\right|_{\rm stat} }\left._{- 0.24}^{+ 0.34}\right|_{\rm syst}\right)\times 10^{-5}\sim \left(2.4\pm 0.5\right)\times 10^{-5}\lb{risu}\eqf} was interpreted by Adam et al. \cite{opera} by
assuming
\eqi v_{\nu_{\mu}}^{\rm meas}=v_{\nu_{\mu}}^{(0)}+\Delta v_{\nu_{\mu}},\eqf with \eqi \Delta v_{\nu_{\mu}}=c\left(\xi_{\rm meas}+1\right) -v_{\nu_{\mu}}^{(0)}>0,\eqf
where $v_{\nu_{\mu}}^{(0)}$ denotes the expected value of the neutrinos' speed: $v_{\nu_{\mu}}^{(0)}\lesssim c$ since the neutrinos have a tiny non-zero mass.
It turned out \cite{opera} that the measured effect, within the accuracy of the measurement, does not depend on the energy of the muon neutrinos in the domain explored by OPERA (some tens of GeV). Such an interpretation of the neutrinic phenomenology observed in OPERA
is unavoidably bound to raise many questions\footnote{In view of the expected forthcoming large amount of papers dealing with \cite{opera}, we will not try to formally cite them here because such a list would likely become out-of-date very quickly. See the electronic databases like ArXiv, NASA/ADS, SPIRES/HEP.} \cite{nature,science,news1,news2,reich}.
\textcolor{black}{However, \cite{cazzoni,cazzoni2} pointed out that the statistically significative positive result of \rfr{risu} might be due to a systematic error of instrumental origin, so that $\xi$ would now be statistically compatible with zero within the error bar.
Recently, a time-of-flight measurement of neutrinos from CERN to LNGS performed by the ICARUS collaboration \cite{contropera}  did not confirm the OPERA result by reporting
\eqi \delta t = 0.3\pm \left.4.0\right|_{\rm stat}\pm \left.9.0\right|_{\rm syst}\ {\rm ns}\sim 0.3\pm 10 \ {\rm ns},\lb{contro}\eqf which is statistically compatible with zero. Note that, as far as the uncertainty in $\xi$ is concerned, \rfr{contro} naively corresponds to about the same level of \rfr{risu}, i.e. about $0.5\times 10^{-5}$; indeed, the accuracies in determining $\delta t$ are about the same in both \rfr{tempis} and \rfr{contro}. }
For earlier studies on superluminal motions of neutrinos in various frameworks, see, e.g., \cite{tachi1,tachi2,tachi3,tachi4,tachi5,tachi6,tachi7}.

Dvali and Vikman \cite{dvali} wondered if the OPERA superluminal phenomenology could be  an environmental effect characteristic of the local neighborhood of the Earth, without the need of violation of the Poincar\'{e} invariance at a fundamental level. Such a scenario, at an effective field theory level, yields naturally  an inevitable appearance of a testable long-range  fifth force of gravitational type.
%
%
%
%
%
Following a clarification by Dvali \cite{dvali2}, it is
%
%
\eqi M_{\ast} \sim \rp{M_{\rm P}}{\xi\sqrt{\alpha}}.\lb{dvaliformula}\eqf In it, $M_{\rm P}\doteq \sqrt{\rp{\hbar c^5}{8\pi G}} = 2.43\times 10^{18}\ {\rm GeV}$ is the energy equivalent of the reduced Planck's mass, where $\hbar$ is the reduced Planck constant, $c$ is the speed of light in vacuum and $G$ is the Newtonian constant of gravitation, and the scale $M_{\ast}$ sets the strength  of the coupling of the putative new massive spin-2 degree of freedom to the neutrino \cite{dvali}.
For other investigations involving various aspects of gravitation, astrophysics and cosmology, see, e.g., \cite{gr1,gr2,gr3,gr4,gr5,gr6,gr7,gr8,gr9,gr10,gr11,gr12,screk}. According to Dvali \cite{dvali2}, the range length $\uplambda$ should not be shorter then the terrestrial radius $ R_{\rm e}$: see also \cite{dvali0}, in which
Earth-size extra dimensions were studies. Data from solar neutrinos, not yet analyzed in this respect, would allow to obtain an upper bound on $\uplambda$ which would likely be shorter than the Earth-Sun distance \cite{dvali2}. Thus, Dvali \cite{dvali2} argues that $\uplambda$ should be
  something less than the Earth-Moon distance, and larger than Earth's radius.

In this paper, we will explore such an intriguing possibility by analytically working out some orbital effects of a gravitational long-range  fifth force of Yukawa-type. We will perform a first-order perturbative calculation without making any a-priori assumptions either on the size of $\uplambda$ or on the trajectory's configuration  of the test particle orbiting the central body acting as source of the putative exotic effect. We will also put constraints on the strength parameter $\upalpha$ for various values of $\uplambda$ within the ranges envisaged by Dvali and Vikman \cite{dvali} in view of the latest results from the orbital determination of some natural and artificial bodies around the Earth. Thus, it will be possible to infer  lower bounds on the coupling of neutrino to the new hypothetical force, which is given by
$M_{\rm P}/M_{\ast}$ \cite{dvali}.
The upper bound on $M_{\rm P}/M_{\ast}$ comes from astrophysical and cosmological observations like star cooling and Big-Bang Nucleosynthesis (BBN) \cite{dvali0}: as\footnote{Dvali and Vikman \cite{dvali} used the reduced Planck units, in which $M_{\rm P} = 1$, and $c=1$.}  Dvali and Vikman  have shown \cite{dvali},  the requirement that new particle is not affecting star-cooling  and BBN yields
$M_{\ast}/M_{\rm P}  >  4\times 10^{-12}-10^{-11}$ or so, \textcolor{black}{corresponding to $M_{\ast}> 10^7-10^8\ {\rm GeV}$ in equivalent energy units}.

The plan of the paper is as follows. In Section \ref{due} we analytically work out some long-term orbital effects due to a Yukawa-like modification of the Newtonian inverse-square law. In Section \ref{tre} and Section \ref{quattro} we phenomenologically constrain  $\upalpha$ with the laser data of the Earth's artificial  satellite LAGEOS II and the Moon, respectively. We also infer corresponding bounds on $M_{\ast}/M_{\rm P}$. Section \ref{cinque} summarizes our findings.
\section{Analytical calculation of some orbital effects induced by a Yukawa-like fifth force}\lb{due}
The Yukawa-type correction to the usual Newtonian gravitational potential $U_{\rm N}=-\mu/r$, where $\mu\doteq GM$ is the gravitational parameter of the central body of mass $M$ which acts as source of the supposedly modified gravitational field,
 is \eqi \Delta U_{\rm Y}=-\rp{\upalpha\mu_{\infty}}{r}\exp\left(-\rp{r}{\uplambda}\right),\lb{uiu}\eqf where $\mu_{\infty}$ is the gravitational parameter evaluated at distances $r$ much larger than the scale length $\uplambda$.
 The total acceleration resulting from
 \eqi U_{\rm tot} = U_{\rm N}+\Delta U_{\rm Y}=-\rp{\mu_{\infty}}{r}\left[1+\upalpha\exp\left(-\rp{r}{\uplambda}\right)\right]\eqf
 is, thus,
 \eqi A_{\rm tot} = -\rp{\mu_{\infty}}{r^2}\left[1+\upalpha\left(1+\rp{r}{\uplambda}\right)\exp\left(-\rp{r}{\uplambda}\right)\right].\lb{accy}\eqf
It should be noticed that  Dvali and Vikman \cite{dvali} leave room, in principle, for a composition-dependent fifth-force, so that $\upalpha$ may not be the same for different bodies.
 From \rfr{accy} the following considerations can be traced about the relation between $\mu_{\infty}$ and the values $\mu_{\rm meas}$ of the gravitational parameter actually measured in, e.g., ranging experiments to terrestrial  artificial and natural satellites, interplanetary probes and planets themselves.
 Indeed, since the Yukawa-like corrections to the Newtonian accelerations felt by the test particles  are usually not included in the dynamical force models fit to the observations, an \virg{effective} value of the gravitational parameter is, actually, measured, i.e. it is
 \eqi \mu_{\rm meas}=\mu_{\infty}\left[1+\upalpha\left(1+\rp{r}{\uplambda}\right)\exp\left(-\rp{r}{\uplambda}\right)\right].\lb{mu}\eqf
 This implies that
 \begin{equation}
\left\{
\begin{array}{lll}
\mu_{\rm meas} & \approx & \mu_{\infty},\ r\gg\uplambda, \\ \\
\mu_{\rm meas}& = & \mu_{\infty}\left[1+2\upalpha\exp\left(-1\right)\right],\ r=\uplambda, \\ \\
\mu_{\rm meas}& \approx & \mu_{\infty}(1+\upalpha), \ r\ll\uplambda.
\end{array}
\right.
\end{equation}
Since $\upalpha$ is, of course, expected to be quite small, it is reasonable and adequate to assume
\eqi\mu_{\infty}\approx \mu_{\rm meas}\eqf also for $r\lesssim\uplambda$
in practical calculations of the perturbative effects of \rfr{uiu} (see \rfr{perihe} below); our further analysis will show a-posteriori that this is just the case, given the upper bounds on $\upalpha$ which will be inferred. Strictly speaking, the use of the measured values $\mu_{\rm meas}$  in those places in the formulas in which $\mu_{\infty}$ appears would be justified only if $\uplambda$ was much smaller than $r$: this would be a fatal restriction because, e.g., $\mu_{\odot}$ is routinely measured from interplanetary ranging mainly involving the inner planets of the solar system, i.e. one would be forced to consider only the case $\uplambda \ll 0.38\ {\rm au}=6\times 10^{10}$ m.

 In view of a first-order perturbative calculation, it is, now, useful to  evaluate $\Delta U_{\rm Y}$ onto the unperturbed Keplerian ellipse and average it over one orbital revolution of the test particle. By using the eccentric anomaly $E$ as fast variable, the result is
\eqi\left\langle \Delta U_{\rm Y}\right\rangle=-\rp{\upalpha \mu_{\infty}\exp\left(-\rp{a}{\uplambda}\right)}{a}I_0\left(\rp{ae}{\uplambda}\right), \lb{poti}\eqf
where $a, e$ are the semimajor axis and the eccentricity, respectively, of the orbit of the test particle, and $I_0(x)$ is the modified Bessel function of the first kind\footnote{See on the WEB http://mathworld.wolfram.com/ModifiedBesselFunctionoftheFirstKind.html and references therein.} $I_k(x)$ for $k=0$.
%
%
Note that \rfr{poti} is an exact result: no approximations have been used either for the orbital configuration of the test particle or the size of the scale parameter $\uplambda$.
%
%
%
%
%
%
%
%
%
%
%
%
%
From \rfr{poti} it is possible to obtain perturbatively the secular precessions of both the pericenter $\omega$ and the mean anomaly ${\mathcal{M}}$ by using the Lagrange planetary equations \cite{BeFa}.

Concerning $\omega$, we have
\eqi \left\langle\dot\omega_{\rm Y}\right\rangle=\upalpha\sqrt{\rp{\mu_{\infty}(1-e^2)}{a}}\rp{\exp\left(-\rp{a}{\uplambda}\right)}{e\uplambda}I_1\left(\rp{ae}{\uplambda}\right),\lb{perihe}\eqf
where $I_1(x)$ is the modified Bessel function of the first kind\footnote{See on the WEB http://mathworld.wolfram.com/ModifiedBesselFunctionoftheFirstKind.html and references therein.} $I_k(x)$ for $k=1$.
%
%
Notice that \rfr{perihe} agrees with  the result obtained by \cite{Bur} with a different approach. More specifically, \cite{Bur} worked out the Yukawa-like pericenter advance per orbit: it can  straightforwardly be obtained from \rfr{perihe} by taking the product of $\left\langle\dot\omega_{\rm Y}\right\rangle$ times the orbital period $P_{\rm b}\doteq 2\pi/n=2\pi\sqrt{a^3/\mu_{\infty}}$. The precession of \rfr{perihe} loses its meaning for $e\rightarrow 0$ since it yields $0/0$.
Other derivations of either the Yukawa-type secular precession of the pericenter or its advance per orbit can be found in, e.g., \cite{Tal,Fisch99,Nord00,Iorio02,Adel,Lucchesi03,Mel04a,Mel04b,Berto,Rey,Ser,Adkins,IorPSS,IorJHEP,IorSYREXE,ijmpdMOF,Cinesi,Hara11a,Hara11b,Hara11c}. All of them make use of different level of approximations in either the magnitude of the length scale $\uplambda$ or the orbital configuration of the test particle.
From \rfr{perihe} it is possible to infer \eqi|\upalpha|\leq \delta(\Delta\dot\varpi)\sqrt{\rp{a}{\mu_{\infty}(1-e^2)}}\rp{e\uplambda\exp\left(\rp{a}{\uplambda}\right)}{I_1\left(\rp{ae}{\uplambda}\right)},\lb{upper}\eqf
where $\delta(\Delta\dot\varpi)$ can be thought as the uncertainty in some observationally determined correction $\Delta\dot\varpi$ to the standard secular precession of the pericenter for some astronomical system. We will use \rfr{upper} in Section \ref{tre} and Section \ref{quattro} for the geodetic satellite LAGEOS II and the Moon, respectively.
\section{Constraints from laser-ranging}
Let us start to consider the range $10^6-10^7$ m corresponding to approximately $1-10$ Earth's radii $ R_{\rm e}$. The traditional constraints on $\upalpha$  for $\uplambda$ lying in the aforementioned range are depicted in Figure 1 of \cite{Kra} or Figure 4 of \cite{Adel}, based on Figure 2.13 of \cite{Fisch99} and adapted by \cite{Mof} in his Figure 4, Figure 1 of \cite{Tal}, and Figure 1 of \cite{Nord}: they are of the order of $10^{-5}-5\times 10^{-8}$. The technique with which they have been obtained is described in detail in \cite{Fisch99}. It is based on the determination of the Earth's gravitational parameter $\mu_{\rm e}$ from laser-ranging measurements at the altitudes of the LAGEOS satellite and the Moon, and on ground-based measurements of the terrestrial gravitational acceleration.

More specifically,
the LAGEOS-Moon constraint of the order of $\approx 5\times 10^{-8}$ is obtained in the following way. First, the ratio
\eqi\eta^{'}_{\rm Y}\doteq 2\left[\rp{r_{\rm L}^2 A(r_{\rm L})-r^2_{\rm M}A(r_{\rm M})}{r_{\rm L}^2 A(r_{\rm L})+r^2_{\rm M}A(r_{\rm M})}\right],\lb{theo}\eqf
where\footnote{Here $r_{\rm L/M}$ is a shorthand for designing alternatively either $r_{\rm L}$ or $r_{\rm M}$.} \eqi A(r_{\rm L/M})=-\rp{\mu_{\infty}}{r_{\rm L/M}^2}\left[1+\upalpha\left(1+\rp{r_{\rm L/M}}{\uplambda}\right)\exp\left(-\rp{r_{\rm L/M}}{\uplambda}\right) \right]\eqf denotes the Newtonian+Yukawa acceleration, to be evaluated at distances $r_{\rm L/M}$, is theoretically computed: \rfr{theo} is, by construction, independent of $\mu_{\infty}$. Then, it is  compared  to the analogous  ratio $\eta^{'}_{\rm N}$ computed for the empirically determined values $\mu_{\rm meas}^{(\rm SLR)}$ and $\mu_{\rm meas}^{(\rm LLR)}$ of the Earth's gravitational parameter at $r_{\rm L}$ and $r_{\rm M}$ from Satellite Laser Ranging (SLR) and Lunar Laser Ranging (LLR) measurements divided by the square of the Earth's radius $ R_{\rm e}$: in the computation of $\eta^{'}_{\rm meas}$ we assume Newtonian values for the accelerations $A(r_{\rm L/M})$. Basically, $\eta^{'}$ is the difference between the values of $\mu_{\rm e}$ evaluated at two different distances normalized to the average  of such two values: in the Newtonian dynamics it vanishes, while in the framework  of the Yukawa-like deviations from the Newtonian picture $\eta^{'}$ is different from zero. The comparison between such two determinations of $\eta^{'}$ yields upper bounds on $\upalpha$ for various values of $\uplambda$.

Instead, the LAGEOS-Earth constraint of about $10^{-5}$ comes from a comparison between the empirical ratio
\eqi \eta_{\rm N} \doteq \rp{A_{\rm terr}( R_{\rm e})-A_{\rm L}( R_{\rm e})}{A_{\rm L}( R_{\rm e})},\eqf in which, as usual, standard Newtonian physics is assumed,
 and the computed one by including the Yukawa term. Notice that $A_{\rm terr}( R_{\rm e})$ denotes the acceleration of gravity measured on the Earth's surface with ground-based techniques, while $A_{\rm L}( R_{\rm e})$ is a shorthand for the ratio of $\mu_{\rm meas}^{(\rm SLR)}$, i.e. the Earth's gravitational parameter empirically determined with the  SLR observations to LAGEOS, to the square of the Earth's radius. When $\eta_{\rm Y}$ is computed, $A_{\rm L}( R_{\rm e})$ is replaced by the product of the Newtonian+Yukawa acceleration $A_{\rm N+Y}(r_{\rm L})$ evaluated at distance $r_{\rm L}$ times $(r_{\rm L}/ R_{\rm e})^2$, while  $A_{\rm terr}( R_{\rm e})$ is replaced by the Newtonian+Yukawa acceleration evaluated on the Earth's surface $A_{\rm N+Y}( R_{\rm e})$. Also $\eta_{\rm Y}$ is independent, by construction, of $\mu_{\infty}$. Instead, \cite{Fisch99} insert $\mu_{\rm meas}^{(\rm SLR)}/R^2_{\rm e}$ in the denominator of $\eta_{\rm Y}$ instead of posing $A_{\rm N+Y}(r_{\rm L})(r_{\rm L}/ R_{\rm e})^2$: this is not consistent with all the previously followed line of reasoning, also because in such a way $\mu_{\infty}$ would not be cancelled in $\eta_{\rm Y}$. Basically, $\eta$ consists of the difference between the accelerations of gravity at the same distance, i.e. on the Earth's surface, measured with different techniques normalized to the value at the same distance obtained with one of such two techniques; again, in Newtonian physics it is expected to be zero, contrary to the Yukawa-type case. Note that in $\eta$ the value obtained from SLR is extrapolated to the Earth's surface by means of the multiplicative scaling factor $(r_{\rm L}/ R_{\rm e})^2$.

 Notice that  both for $\eta^{'}$ and $\eta$ the scale length $\uplambda$ has been kept fixed.
\subsection{The perigee of LAGEOS II}\lb{tre}
Actually, tighter constraints can be obtained by using the perigee of\footnote{For preliminary investigations on the possibility of using the perigee of LAGEOS II to constrain  Yukawa-like deviations from the Newtonian inverse square law of gravity, see \cite{Iorio02,Lucchesi03}. For earlier investigations on the effect of a Yukawa-like fifth force on the perigee of LAGEOS-type satellites, see \cite{Nord98}.} LAGEOS II in connection with \rfr{perihe} and \rfr{upper}. The semi-major axis of LAGEOS II is $a=12,163\ {\rm km}= 1.2163\times 10^{7}\ {\rm m}=1.9\  R_{\rm e}$, while its eccentricity is $e=0.014$.

It is well known that the perigees of the LAGEOS-type satellites are particularly sensitive to a host of non-gravitational perturbations which are, thus,  a major limiting factor in constraining $\alpha$.
 As pointed out by Ries et al. \cite{Ries03} in the context of the re-analysis of the earlier tests of the general relativistic Lense-Thirring effect with the LAGEOS and LAGEOS II satellites \cite{Ciu98}, the realistic accuracy in determining the secular perigee precession of LAGEOS II is larger than the gravitomagnetic effect itself mainly because of the non-gravitational perturbations affecting such an orbital element. From Table 9.7 of \cite{Lucchesi} it is possible to infer  an uncertainty of about 125 milliarcseconds per year (mas yr$^{-1}$ in the following).

Another important source of systematic uncertainty is the mismodeling in the multipoles of the expansion of the non-spherical  part of the gravitational potential. In particular, the fully normalized even ($ \ell = 2,4,\ldots $) zonal ($ m = 0 $) Stokes coefficients $\overline{C}_{\ell,0}$ cause secular perigee precessions which acts as superimposed bias. Assessing correctly their mismodeling is crucial for a realistic evaluation of the overall uncertainty affecting $\alpha$.

There are nowadays several institutions worldwide   which almost routinely produce global gravity field solutions by analyzing increasing data sets from the dedicated spacecraft-based missions like CHAMP\footnote{It re-entered Earth's atmosphere on September 20, 2010.}, GRACE, GOCE, along with observations collected from many geodetic satellites of LAGEOS-type tracked by the International Laser Ranging Service \cite{SLR}. Among their products,  collected by the International Centre for Global Earth Models (ICGEM) \cite{icgem}, the  coefficients $\overline{C}_{\ell,m},\overline{S}_{\ell,m}$ of the  geopotential stand out.

For our purposes, it is fundamental to realistically evaluate the uncertainties in the zonals, with particular regard to $\overline{C}_{2,0}$ since it induces the largest aliasing secular perigee precession.
To this aim, we will follow the approach recently described by Wagner and McAdoo \cite{Wag012}.
They felt the need of comparing not only solutions releasing the mere statistical, formal errors for the geopotential coefficients, but also older models yielding calibrated errors, by explicitly requiring that the benchmark model must be formally far more accurate than the one to be tested.  Furthermore, Wagner and McAdoo \cite{Wag012} pointed out that their method could well be applied even to solutions not displaying formal errors at all.
Wagner and McAdoo \cite{Wag012} remarked that the calibration of the errors in a given test model should be made by using reference solutions obtained independently: for example, a GRACE-based solution should be compared with, say, a GOCE-based solution, as done by Wagner and MacAdoo  themselves \cite{Wag012}. Even in such a case, care should be taken to avoid that the reference model adopted was not used as a-priori background model in producing the models to be tested \cite{Wag012}.

Here we apply the method by Wagner and McAdoo \cite{Wag012} by choosing the CHAMP-based solution AIUB-CHAMP03S \cite{aiubchamp03s}, the GRACE/GOCE/CHAMP/SLR-based solution GOCO2S \cite{goco2s}  and the GOCE-only solution GOCONS \cite{gocons} as test models, while we take the wholly independent GRACE/GOCE/LAGEOS-based solution EIGEN-6 \cite{eigen6c} as formally superior reference model. EIGEN-6 was released by the GeoForschungsZentrum (GFZ),  AIUB-CHAMP03S was produced by the Astronomisches Institut,  Universit\"{a}t Bern (AIUB), while GOCO2S and GOCONS were released by the Gravity Observation COmbination (GOCO) consortium. The background gravity models adopted for AIUB-CHAMP03S were the pre-CHAMP/GRACE/GOCE JGM3 \cite{JGM3} and EGM96 \cite{EGM96} SLR-only solutions, while  EIGEN-6 used 6.5 years of LAGEOS  data from the time span 1 Jan 2003 till 30 June 2009. GOCO2S used the GRACE-only solution ITG-Grace2010s \cite{itg} as background model. GOCONS did not adopt any background  model, and its data from GOCE cover the period 1 November 2009-17 April 2011. See the discussion in \cite{Wag012} about the risk of precluding an unbiased calibration employing an external standard model.
As requested by Wagner and McAdoo \cite{Wag012}, the formal $\sigma_{\overline{C}_{2,0}}$ of EIGEN-6 is smaller than those of the models to be tested by about 1 order of magnitude.
\begin{table*}[ht!]
\caption{First row: uncertainty $\delta\overline{C}_{2,0}$ in the even zonal of degree $\ell=2$ evaluated according to the method by Wagner and McAdoo \cite{Wag012}. EIGEN-6 \cite{eigen6c} was assumed as formally superior reference model ($\sigma_{\overline{C}_{2,0}} = 2\times 10^{-13}$), while the solutions  AIUB-CHAMP03S \cite{aiubchamp03s}, GOCONS \cite{gocons}, and GOCO2S \cite{goco2s} were taken as test models.  The tide system of all the models considered is tide-free. Second row: formal  $\sigma_{\overline{C}_{2,0}}$ of the aforementioned models, and calibrated $\sigma_{\overline{C}_{2,0}}$ of GIF48, whose tide system is zero-tide.
}\label{geotavola}
\centering
\bigskip
\begin{tabular}{lllll}
\hline\noalign{\smallskip}
 & AIUB-CHAMP03S (test) & GOCONS (test) & GOCO2S (test) & GIF48 \\
\noalign{\smallskip}\hline\noalign{\smallskip}
$\delta\overline{C}_{2,0}$  & $3.3\times 10^{-11}$ & $1.09\times 10^{-10}$ & $1.08\times 10^{-10}$  & - \\
$\sigma_{\overline{C}_{2,0}}$& $1.1\times 10^{-11}$ & $1.9\times 10^{-11}$ & $4\times 10^{-13}$ & $7.0\times 10^{-11}$ \\
\noalign{\smallskip}\hline\noalign{\smallskip}
\end{tabular}
\end{table*}
Instead of using a single coefficient error scaling factor for the formal sigmas $\sigma_{\overline{C}^{\rm test}_{\ell,0}}$ of each of the test models, Wagner and McAdoo \cite{Wag012} also proposed to average the individual error factors over all the $2\ell +1$ coefficients of degree $\ell$. According to eq. (A13) of \cite{Wag012}, one has
\eqi \overline{f}_{\rm test, \ell}=\left\{\left(\rp{1}{2\ell +1}\right)\sum_{m=0}^{2\ell+1}\left[\rp{\left(\overline{H}_{\ell,m}^{\rm test}-\overline{H}_{\ell,m}^{\rm ref}\right)^2-\left(\sigma^{\rm ref}_{\overline{H}_{\ell,m}}\right)^2}{\left(\sigma^{\rm test}_{\overline{H}_{\ell,m}}\right)^2}\right]\right\}^{\rp{1}{2}},\lb{A13}\eqf
where $\overline{H}_{\ell,m}$ denotes both $\overline{C}_{\ell,m}$  and $\overline{S}_{\ell,m}$ in the sense that the sum in \rfr{A13} includes all the geopotential coefficients of both kinds for a given degree $\ell$. Our results are displayed in Table \ref{geotavola}. It can be noticed that the resulting uncertainties $\delta{\overline{C}_{\ell,0}}$ are up to $1-3$ orders of magnitude larger than the formal errors. Incidentally, they are rather close to the calibrated error released in the GRACE-only GIF48 model  by CSR ($\delta{\overline{C}_{2,0}}=7.0\times 10^{-11}$).

We will adopt our results in Table \ref{geotavola} to assess the zonals-induced bias on the Yukawa-induced perigee orbital precession of LAGEOS II.
Thus, by using such a figure in \rfr{upper}, it is possible to plot the upper bound on $\upalpha$ as a function of $\uplambda$: see Figure \ref{lageos2}.
\begin{figure}
\begin{center}
\includegraphics[width=\columnwidth]{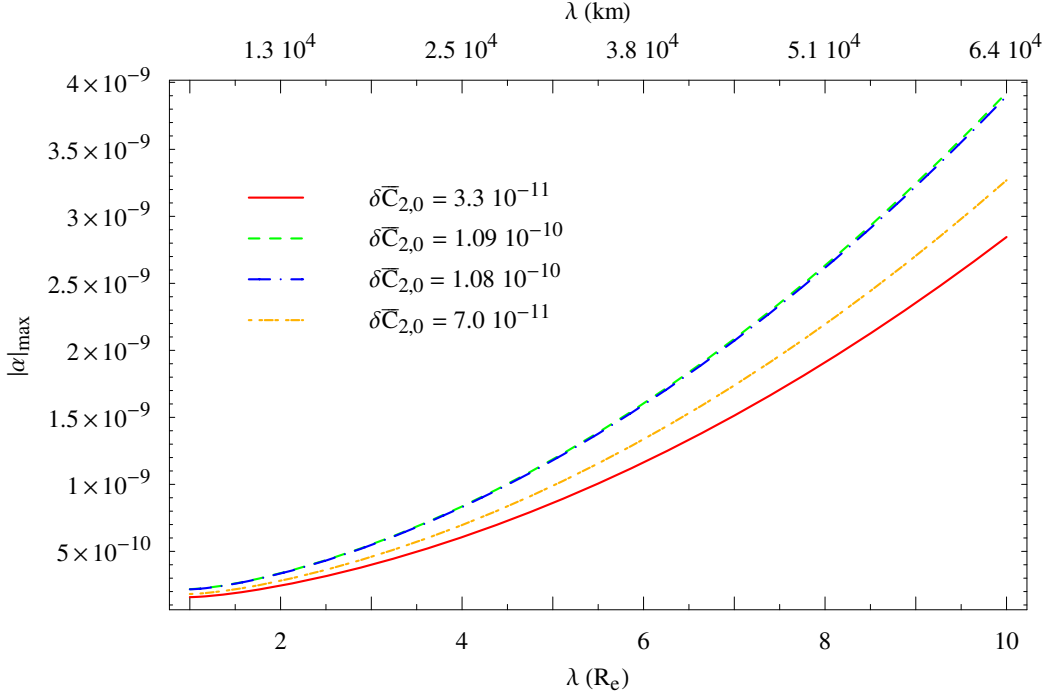}
 \caption{Upper bound on $\left|\upalpha\right|$ for the Earth-LAGEOS II system from the uncertainty in  the precession of the perigee $\omega$ of LAGEOS II as a function of $\uplambda$  for $1\  R_{\rm e}\leq\uplambda\leq 10$ $ R_{\rm e}$. As far as the mismodeling $\delta\overline{C}_{2,0}$ in the first even zonal harmonic of the Earth's geopotential is concerned, we adopted the figures of Table \ref{geotavola} for it, while the uncertainty due to the non-gravitational perturbations was taken as large as 125 mas yr$^{-1}$ \cite{Lucchesi}.}\lb{lageos2}
 \end{center}
\end{figure}
It can be noted that there is an improvement of $2-5$ orders of magnitude with respect to Figure 1 of \cite{Kra}, Figure 1 of \cite{Tal}, and Figure 4 of \cite{Adel}; see also Figure 1 of \cite{Nord}.
%
%
%
%
%
%
%
%
%
%
Lucchesi and Peron \cite{Lucc010}, using the approximate analytical results by \cite{Lucchesi03} and a data record 13 yr long for LAGEOS II, claim $\left|\upalpha\right|\leq 9.9\times 10^{-12}$ at $\uplambda=1\  R_{\rm e}$. However, such a figure should be considered just as a preliminary result coming from the statistical errors of the linear  fitting of the post-fit numerically integrated residuals of the perigee since, as pointed out by  Lucchesi and Peron themselves \cite{Lucc010}, it does not include an analysis of the systematic errors due to the uncertainties in the zonals' geopotential. Lucchesi and Peron \cite{Lucc010} used the GRACE-only model EIGEN-GRACE02S \cite{eigengrace02s}, which released a calibrated $\sigma_{\overline{C}_{2,0}} = 5.304 \times 10^{-11}$. Moreover, Lucchesi and Peron \cite{Lucc010} did not explicitly include a Yukawa-like fifth force in the dynamical models fitted to the data of LAGEOS II, by assuming that it would entirely be accounted for by their numerically integrated residuals of the perigee. A similar approach with the LAGEOS satellite was followed by March et al. \cite{march} in constraining other non-Newtonian putative effects.

The bounds on $\alpha$ of Figure \ref{lageos2} yield corresponding constraints on $M_{\ast}/M_{\rm P}$ according to \rfr{dvaliformula}. They are depicted in Figure \ref{dvaliplot}.
\begin{figure}
\begin{center}
\includegraphics[width=\columnwidth]{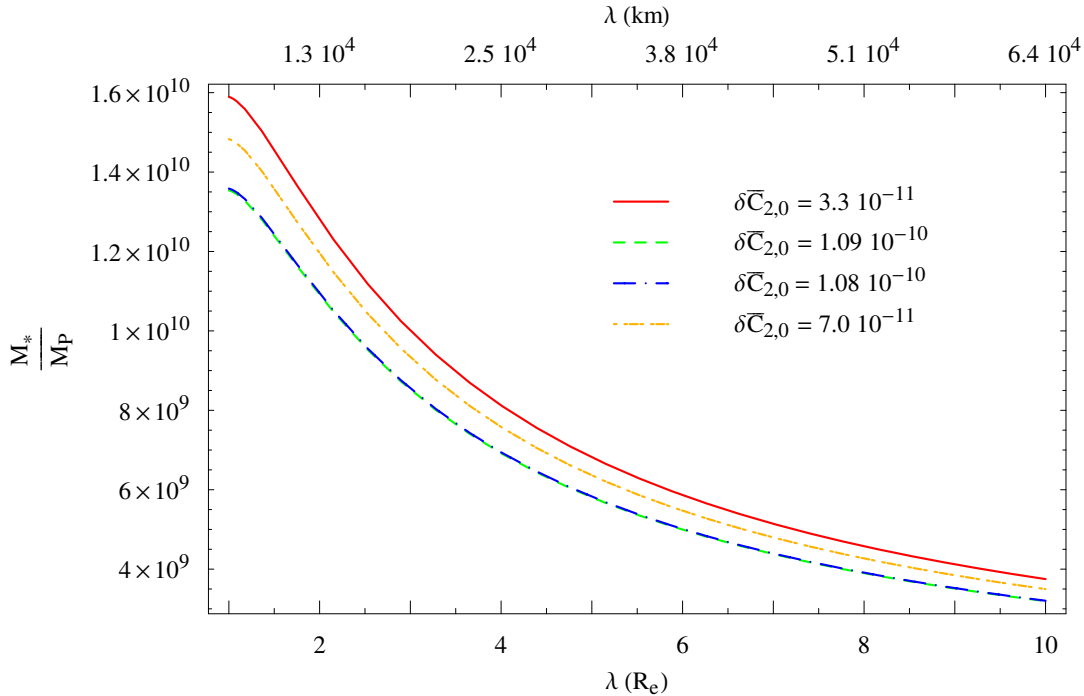}
 \caption{\textcolor{black}{Lower limits on} $M_{\ast}/M_{\rm P}$ from \rfr{dvaliformula} for $\xi \textcolor{black}{\lesssim} 0.5 \times 10^{-5}$ \cite{opera,contropera} and the bounds on $\alpha$ of Figure \ref{lageos2}. }\lb{dvaliplot}
\end{center}
\end{figure}
For $\uplambda = 1-1.5\  R_{\rm e}$, it is $M_{\ast}/M_{\rm P}\textcolor{black}{\gtrsim} 1.4-1.6\times 10^{10}$, while for $\uplambda = 10\  R_{\rm e}$ we have $M_{\ast}/M_{\rm P}\textcolor{black}{\gtrsim} 4-5\times 10^{9}$. \textcolor{black}{Such results are in striking contrast with $M_{\ast}\sim 10^{-6}M_{\rm P}$ inferred by Dvali and Vikman \cite{dvali} for $\uplambda = 1\ R_{\rm e}$.}
\subsection{The perigee of the Moon}\lb{quattro}
Let us, now, consider the range $10-100$ $ R_{\rm e}$ corresponding to $6.4\times 10^7-6.4\times 10^8$ m; the lunar semi-major axis is $a_{\rm M}=60\  R_{\rm e}=3.8\times 10^8$ m. To this aim, the motion of the Moon is best suited to yield tight constraints.

Reasoning in term of the lunar perigee, the uncertainty in determining its secular precession is of the order of $\delta\dot\varpi\approx 0.1$ mas yr$^{-1}$ \cite{Dick,Muller1,Williams,Muller2,Muller3}.
The resulting bounds on $\upalpha$ are depicted in Figure \ref{luna}.
\begin{figure}
\begin{center}
\includegraphics[width=\columnwidth]{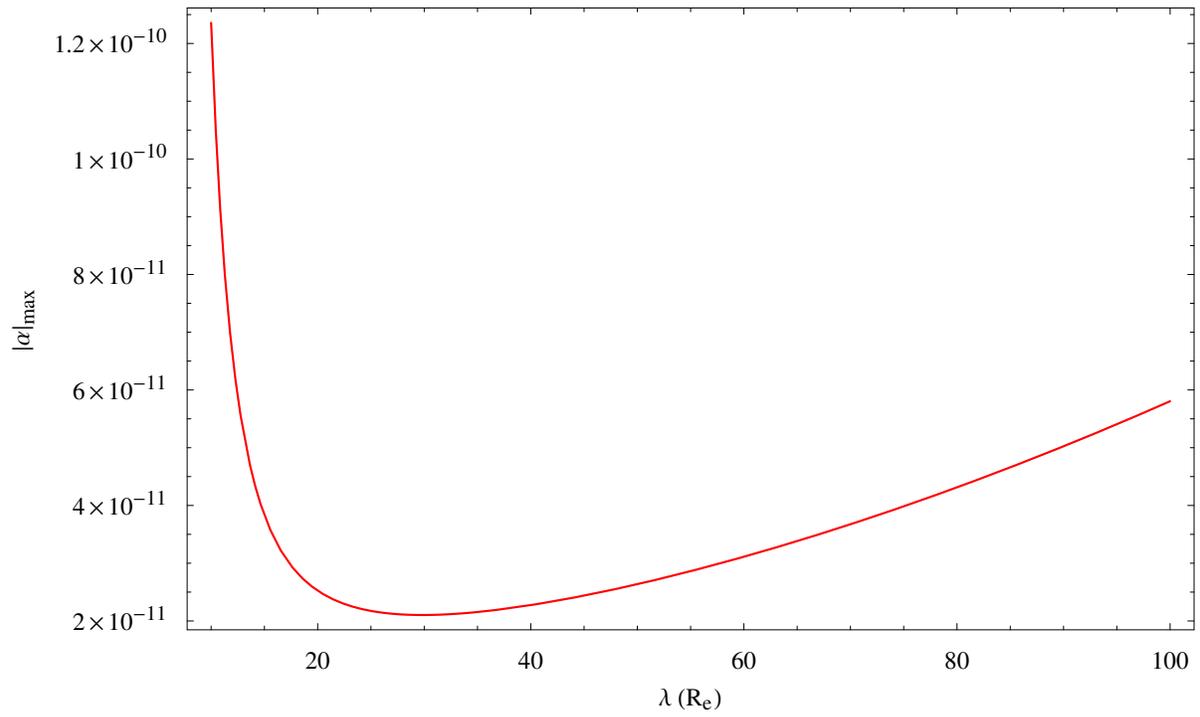}
 \caption{Upper bound on $\upalpha$ for the Earth-Moon system from the uncertainty in determining the precession of the perigee of the Moon, assumed as large as 0.1 mas yr$^{-1}$ \protect\cite{Dick,Muller1,Williams,Muller2,Muller3}, as a function of $\uplambda$  for $10\  R_{\rm e}\leq\uplambda\leq 100$ $ R_{\rm e}$.}\lb{luna}
\end{center}
\end{figure}
It can be noted that $|\upalpha|\lesssim 3\times 10^{-11}$ for $\uplambda=60  R_{\rm e}= 3.8\times 10^8$ m, in substantial agreement with Figure 1 of \cite{Nord}, Figure 1 of \cite{Kra} and Figure 4 of \cite{Adel}. Cfr. also with the upper bound of $\left|\upalpha\right|\leq 5\times 10^{-11}$ ($\uplambda=60\ R_{\rm e}$) by\footnote{M\"{u}ller et al. \cite{Muller2,Muller3} actually included a Yukawa-type fifth force in the mathematical force models with which they analyzed the LLR data.} M\"{u}ller et al. \cite{Muller2,Muller3}. However,  the constraints of Figure \ref{luna} are orders of magnitude better than those reported
in \cite{Nord,Kra,Adel} for $\uplambda \neq 60  R_{\rm e}$. We also note that\footnote{I thank W.-T. Ni for having pointed out it to me.} \cite{boh} obtained $\left|\upalpha\right|\lesssim 10^{-8}$ for $1.8\  R_{\rm e}\lesssim \uplambda \lesssim 60\  R_{\rm e}$.

In Figure \ref{dvaliplotluna} we display the corresponding bounds on $M_{\ast}/M_{\rm P}$.
\begin{figure}
\begin{center}
\includegraphics[width=\columnwidth]{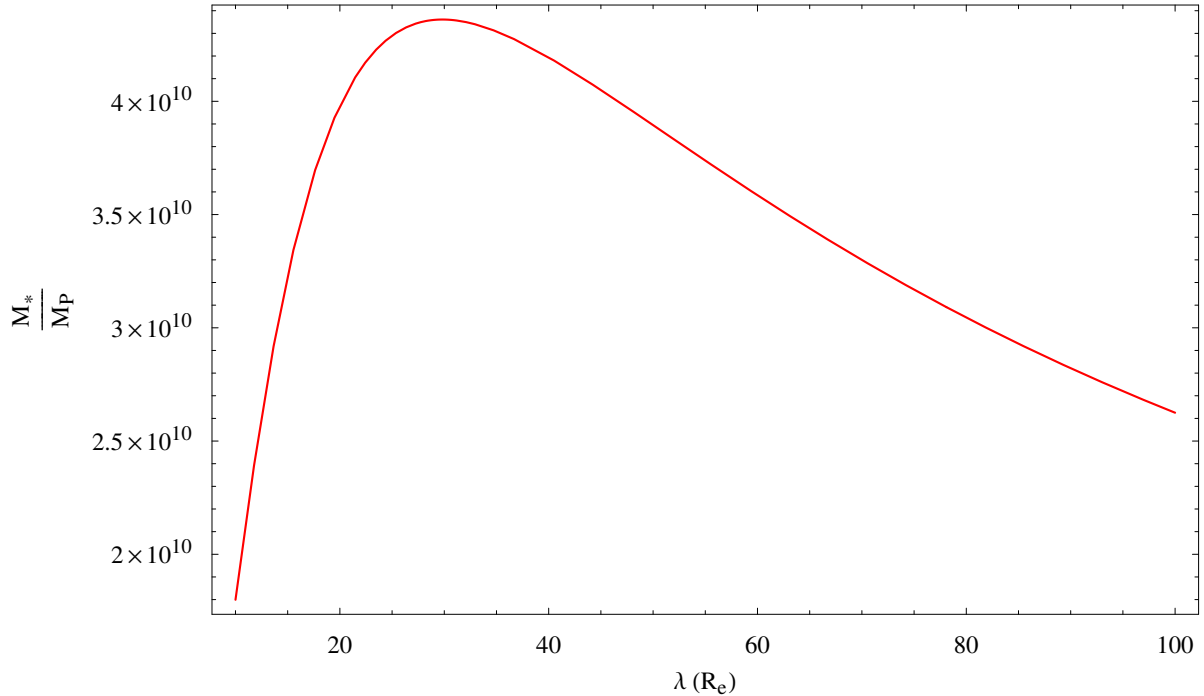}
 \caption{\textcolor{black}{Lower bounds on} $M_{\ast}/M_{\rm P}$ from \rfr{dvaliformula} for $\xi\textcolor{black}{\lesssim} 0.5 \times 10^{-5}$ \cite{opera} and the lunar bounds on $\alpha$ of Figure \ref{luna}. }\lb{dvaliplotluna}
 \end{center}
\end{figure}
While for $\uplambda =10\  R_{\rm e}$ it is $M_{\ast}/M_{\rm P}\sim 2\times 10^{10}$, we have $M_{\ast}/M_{\rm P}\textcolor{black}{\gtrsim} 3.5\times 10^{10}$ at $\uplambda\sim 60\ R_{\rm e}$, with  $M_{\ast}/M_{\rm P}\textcolor{black}{\gtrsim} 4.5\times 10^{10}$ at $\uplambda \sim 30\  R_{\rm e}$.
%
%
%
%
%
%
%
%

Finally, we remark that using  data from the Lunar Reconnaissance Orbiter (LRO) should allow to further improve the bounds on $\left|\upalpha\right|$ at $\uplambda = 60\  R_{\rm e}$ \cite{Tsang}.

\section{Discussions and conclusions}\lb{cinque}
We analytically worked out some orbital effects induced by a hypothetical gravitational fifth force of Yukawa-type  on the orbital motion of a test particle moving around a central body acting as source of the modified gravitational field. We  restricted ourselves neither to any specific size for the scale length parameter $\uplambda$ nor to small values of the orbital eccentricity $e$ of the particle. We obtained secular precessions for  the pericenter $\omega$. Our results imply the use of the modified Bessel functions  of the first kind $I_k(x),\ k=0,1$.

We used the Satellite Laser Ranging data of the artificial satellite LAGEOS II to put constraints on the coupling strength parameter $\upalpha$ of the putative fifth force for the Earth-LAGEOS II system in the range $1\  R_{\rm e}\leq \uplambda\leq 10\  R_{\rm e}$ obtaining $1\times 10^{-10}\lesssim \left|\upalpha\right|_{\rm max}\lesssim 4\times 10^{-9}$. The accuracy in determining the Moon's perigee from the Lunar Laser Ranging technique allowed us to infer $\left|\upalpha\right|_{\rm max}\sim 3\times 10^{-11}$ for the Earth-Moon pair in the range $15\  R_{\rm e}\lesssim \uplambda\lesssim 60\  R_{\rm e}$.
\textcolor{black}{From $M_{\ast}/M_{\rm P}\sim 1/\xi\sqrt{\alpha}$ and by using $\xi \lesssim 0.5\times 10^{-5}$, coming from the overall uncertainty in the neutrinic time-of-flight measurements common to both OPERA and ICARUS collaborations, our bounds on $\alpha$} correspond to $M_{\ast}/M_{\rm P}\textcolor{black}{\gtrsim} 4\times 10^9 - 1.6\times 10^{10}$ for $1\  R_{\rm e}\leq \uplambda\leq 10\  R_{\rm e}$, and to $M_{\ast}/M_{\rm P} \textcolor{black}{\gtrsim} 3\times 10^{10}-4.5\times 10^{10}$ for $15\  R_{\rm e}\lesssim \uplambda\lesssim 60\  R_{\rm e}$, with the maximum value occurring at $\uplambda\sim 30\  R_{\rm e}$. \textcolor{black}{Such bounds on $M_{\ast}$ disagree with $M_{\ast}\sim 10^{-6}M_{\rm P}$ by Dvali and Vikman}.

Our results are \textcolor{black}{neither} necessarily limited to the OPERA neutrino scenario \textcolor{black}{nor to the Dvali-Vikman model}, being valid for any theoretical scenario yielding an effective long-range Yukawa-type correction to the Newtonian inverse-square law.

As a final, \textcolor{black}{technical} remark, we notice that a complementary treatment would require that a Yukawa-type extra-force should be explicitly modeled in the softwares used to process the data records, and a dedicated solve-or parameter should be estimated in dedicated data reduction procedures in which the ad-hoc modified dynamical models are fitted to the existing  data sets. Such an approach has not yet been implemented in some of the analyses previously described like the LAGEOS II one: it is a drawback common to the LAGEOS-based Lense-Thirring analyses implemented so far.
\section*{Acknowledgments}
I thank G. Dvali, R. Konoplya and W.-T. Ni for insightful correspondence. I am also grateful to M. Schreck for having disclosed some typos.
\bibliography{Operabib}{}

\end{document}